\documentclass[a4,10pt]{article}
\usepackage[backend=bibtex]{biblatex}

\usepackage{arxiv}
\usepackage{fancyhdr}
\usepackage[T1]{fontenc}    
\usepackage[utf8]{inputenc}
\usepackage{url}            
 \usepackage{cmbright}  

\usepackage{nicefrac}       
\usepackage{microtype}      
\usepackage{lipsum}
\usepackage{graphicx}
\graphicspath{ {./images/} }
\usepackage{tikz}   
\usepackage{enumitem}
\usepackage{amsmath,amssymb,amsfonts}
\usepackage{xcolor}
\definecolor{myorange}{RGB}{100, 50, 0}
\definecolor{myblub}{RGB}{34, 52, 168}
\definecolor{dg}{RGB}{64,64,64}

\usepackage[colorlinks = true,
		citecolor = blue,
		urlcolor= blue,
		linkcolor=dg,]{hyperref}
\usepackage{algorithmic}
\usepackage{textcomp}
\usepackage{xcolor}
\def\BibTeX{{\rm B\kern-.05em{\sc i\kern-.025em b}\kern-.08em
    T\kern-.1667em\lower.7ex\hbox{E}\kern-.125emX}}
    
\usepackage{placeins}
\usepackage{subcaption}
\fancyheadoffset{0pt}
\title{Aggregating Digital Identities through Bridging. An Integration of Open Authentication Protocols for Web3 Identifiers.}

\author{
Ben Biedermann\\
\href{mailto:bb@acurraent.com}{bb@acurraent.com}\\
\textit{ACURRAENT UG};\\
Islands and Small States Institute \\
University of Malta\\
\href{https://orcid.org/0000-0003-1331-6517}{0000-0003-1331-6517}\\
\And
Matthew Scerri\\
\href{mailto:matthew.scerri@gmail.com}{matthew.scerri@gmail.com}\\
\textit{WIDE Consortium}\\
\.Zabbar\\
Malta \\
\And
Victoria Kozlova\\
\href{mailto:vk@acurraent.com}{vk@acurraent.com}\\
\textit{ACURRAENT UG}\\
Frankfurt (Oder) \\
Germany\\
\And
Joshua Ellul\\
\href{mailto:joshua.ellul@um.edu.mt}{joshua.ellul@um.edu.mt}\\
Centre for DLT\\
University of Malta\\
\href{https://orcid.org/0000-0002-4796-5665}{0000-0002-4796-5665}\\
}

\begin{document}

 \pagestyle{plain}
\pagenumbering{arabic}
\maketitle
\pagestyle{fancy}

\begin{abstract}
Web3’s decentralised infrastructure has upended the standardised approach to digital identity established by protocols like OpenID Connect. Web2 and Web3 currently operate in silos, with Web2 leveraging selective disclosure JSON web tokens (SD-JWTs) and Web3 dApps being reliant on on-chain data and sometimes clinging to centralised system data. This fragmentation hinders user experience and the interconnectedness of the digital world. This paper explores the integration of Web3 within the OpenID Connect framework, scrutinising established authentication protocols for their adaptability to decentralised identities. The research examines the interplay between OpenID Connect and decentralised identity concepts, the limitations of existing protocols like OpenID Connect for verifiable credential issuance, OpenID Connect framework for verifiable presentations, and self-issued OpenID provider. As a result, a novel privacy-preserving digital identity bridge is proposed, which aims to answer the research question of whether authentication protocols should inherently support Web3 functionalities and the mechanisms for their integration. Through a Decentralised Autonomous Organisation (DAO) use case, the findings indicate that a privacy-centric bridge can mitigate existing fragmentation by aggregating different identities to provide a better user experience. While the digital identity bridge demonstrates a possible approach to harmonise digital identity across platforms for their use in Web3, the bridging is unidirectional and limits root trust of credentials. The bridge’s dependence on centralised systems may further fuel the debate on (de-)centralised identities.
\end{abstract}



\smallskip
\noindent \textit{This work is licensed under a Creative Commons Attribution-NonCommercial-ShareAlike 4.0 International License.}
 \begin{tikzpicture}[remember picture, overlay]
    \node[anchor=south east, xshift=-0.5cm, yshift=0.5cm] at (current page.south east) {
        \includegraphics[width=2cm]{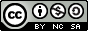} 
    };
\end{tikzpicture}

\section{Introduction}\label{sec_1}
The rise of Web3, a global value network built on decentralised infrastructure, has fundamentally challenged how we think about digital identity. Yet, large initiatives in the digital identity sector have resorted to established identity and authentication protocols. For example, the developers of the European Digital Identity Wallet (EUDIW) have resorted to protocols under OpenID Connect (OIDC)~\cite{kounis_2023}. Once OIDC unified the Web in providing a widely accepted method for handling client-based authentication, OIDC was even referred to as a decentralised identity solution~\cite{weitzner_2007}. Now, the Web finds itself split between applications that continue to use OIDC and ``decentralised'' applications (dApps), which either resorted to storing credential data on-chain or using centralised credentialing solutions. Digital identity in its current form, thus is fragmented, convolutes terms, and lacks interoperability.\\

In part, poor performance of digital identities in covering use cases across traditional and decentralised applications stems from a lack of solutions that attempt to connect Web2 to Web3~\cite{ibrahimy_2022}. Following the evolution of OIDC, when OAuth first was proposed, before blockchain technology was available to be used as a trust layer for asymmetric key pairs, trust management had to rely on public key infrastructure (PKI). Although PKIs still dominate the market, blockchains are now available as an alternative. This circumstance made the client-based execution of OAuth comparably more decentralised than approaches that used public-private key pairs, which had to be stored centrally. It was not until the emergence of programmable blockchains with Ethereum in 2014/15, when asymmetric cryptography started to be perceived as a decentralised source of trust for digital identity solutions and SSI specifically~\cite{schaffner_2019}. The emergence of blockchains prompted experimentation to use the new technology for decentralised PKIs (DPKI). In turn, the term ``\textit{decentralised identity}'' was increasingly used to denote blockchain-based identity solutions rather than digital identity technology using OIDC and OIDC started to evolve into being referred to as ``federated identity'' protocol~\cite{radha_2012}. Following this, in the debate on digital identity solution the use of blockchain became the delineating factor for distinguishing between ``decentralised'' Web3 technologies and ``centralised'' identities in Web2.\\

Technologies across all sub-fields of digital identity fall short of delivering clear and cross-platform solutions for users that need to interact equally with infrastructure in Web2 and Web3. These needs can be as simple as tying an Ethereum wallet address to a Google Workspace account for connecting a payment to the work completed. Until 2027, the European Digital Identity (EUDI) will be issued under the European Commission's \textit{Proposal for a Regulation Amending Regulation (EU) No 910/2014 as Regards Establishing a Framework for a European Digital Identity}, also known as the updated Regulation on electronic identification, authentication, and trust services (eIDAS 2.0)~\cite{european_commission_2021}. However, it requires users to adopt yet another digital identity tool for interacting with third parties, particularly in governmental contexts. Limitations of such a digital identity framework limits access to Web3 use-cases, e.g. users will likely not be able to use their verified university diploma for joining a decentralised autonomous organisation (DAO). Furthermore, SSI wallets that allow users to access some Web3 functionality require users to manage a second cryptographic key pair, connect both their Web3 wallet and their self-sovereign identity (SSI) wallet, and only allow for credential management on edge devices~\cite{reed_2021a}. Even for practitioners in Web3, these steps pose significant hurdles for making use of any additional functionality.\\ 

This work is intended to foster a unified digital identity landscape to the benefit of users and encourage digital identity practitioners to collaborate across technological divides for creating a future for authentication and a decentralised web. To this end, this paper investigates the intersection of Web3-targeted identity solutions and established OIDC protocols that incorporate decentralised identity specifications, such as the verifiable credential (VC) data model~\cite{sporny_2023}. This research proposes a wallet-like architecture that aggregates, encrypts, and attests to credentials from various issuers by integrating a multitude of digital identity protocols. It aims to answer the question: \textit{How can user identities of centralised provenance be efficiently provisioned for activities in Web3 without additional wallet key and credential management on an edge device?} By answering this question, this work fills the gap of lacking use cases for EUDI outside of the prescribed set of pilots, enhances interoperability from OAuth to on-chain attribute attestations, and introduces a novel approach to privacy-preserving credential management on untrusted servers.\\

The rest of the paper is structured as follows. First, the historical relationship between OIDC and the concept of decentralised identity is explained in detail, connecting disparate and overlapping data formats for digital identity and their transport protocols. The paper then outlines the challenges of integrating Web3 with OIDC and the EUDIW, highlighting the limitations of the protocols that are used, such as the latest generation of OIDC for verifiable credential issuance (OID4VCI) and verifiable presentation (OID4VP) with a self-issued OpenID provider (SIOP). In Section~\ref{sec_3}, the proposed architecture and its technical specifications are described. Thereafter, in Section~\ref{sec_4}, the architecture is put into context of use cases that are piloted by the consortium \textit{Web3 Identity for DAOs and Education} (WIDE). Then, the findings and results from sections~\ref{sec_3} and~\ref{sec_4} are discussed considering the trends in digital identity. Finally, conclusions and future research are outlined in Section~\ref{sec_5} to critically engage with the identity bridge as privacy-preserving to disjointed digital identity technologies and initiatives.

\section{Background}\label{sec_2}
This section is divided into two parts. First, subsection~\ref{sec_2.1} introduces and defines key concepts and terms required for evaluating the technical merits of the solution, including self-sovereign identity, sybil resistance, decentralised autonomous organisations (DAOs), and Web3 architectures. Later on, the paper proceeds to position its contribution in the digital identity discourse, attributing the proposal made herein to limitations and contestations of the state-of-the-art in digital identity.

\subsection{Terms of Reference}\label{sec_2.1}
Considering DAOs as a use case for a digital identity bridge that is fundamental for conceiving the concept of bridging identities, it is important to introduce foundational terms related to the application of blockchains. This allows the reader to evaluate and understand the merits of an identity bridging solution to function as a connector and middleware between technology paradigms. To that end, this subsection introduces the terms of reference and definitions for the most important concepts used in this contribution, as well as discusses them in light of different interpretations of \textit{decentralisation}. When referring to the practical adoption of blockchain technology, applications and relationships using blockchain technology are commonly referred to as Web3. According to Wan et al. \textit{Web3 adopts blockchain and digital wallets, using distributed networks to ``empower users to take ownership''}~\cite{wan_2023}. Thus, it is established against the backdrop of the platform economies of Web2~\cite{wan_2023}.\\

Web3 identifiers in this regard are public key pairs to which ownership of assets on blockchains is cryptographically linked. Therefore, they are distinct from Web2 identifiers, which are considered to be usernames and email addresses~\cite{wan_2023}. Instead of being linked to passwords in a centralised database, Web3 identifiers are being held in wallets. These wallets currently offer less functionality than identity wallets, but give the users more freedom to import, export, and add asymmetric key pairs. The standard for this type of key management was first described by the Bitcoin Improvement Proposal (BIP) 32~\cite{wuille_2012}. Ever since, ``\textit{deterministic hierarchical wallets}'' offer liberal functionalities that are focussed on handling cryptographic key pairs for interacting with blockchains. Popular examples for such Web3 wallets are MetaMask, Phantom, and Coinbase Wallet. In 2016, Allen introduced the concept of SSI~\cite{allen_2016}, which would expand the concept of Web3 wallets to applications holding all kinds of credentials, proofs, and personal information~\cite{wang_2020}. SSI was meant to take the learnings of user ownership and user control from Web3 and apply it to applications that were deeply rooted in Web2, such as social media platforms and e-government portals. The term SSI is vague, but sets out ten principles for a new wave of digital identity solutions. By design, the concept is prescriptive, i.e. describes what should be and has caused contestation and confusion among practitioners~\cite{allen_2016,weigl_2022}.\\

Although SSI was found to not be underpinned by a single definition, the academic discourse converges to state that SSI preserves or enables user rights~\cite{wang_2020,kondova_2020}. Hence, it is a functionality-driven definition that also impacts its interplay both with centralised identity and decentralised identity systems. For this reason, in Web3, the notion of decentralisation is reduced to the concept of ownership, which is exercised through the control of the key pair that grants access to the assets recorded on the blockchain~\cite{wang_2020}. Conversely, proponents of OIDC ascribe OAuth-based systems the attribute of decentralisation because of the decentralised control over the reference to the data~\cite{weitzner_2007}.\\

There is also a chasm between different digital identity paradigms in defining decentralisation. While OIDC emphasises control over the data reference to denote decentralisation, Web3 uses the place of storage for defining decentralisation. However, both approaches acknowledge the necessity of a data reference, such as a uniform resource identifier (URI) or public key. This highlights another difference that is the ascribed importance of identification. OIDC emerged as an authentication framework, which was retrofitted to allow for identification across platforms~\cite{miller_2007,russell_2007}, whereas Web3 focused on authorisation and only recently allowed for both authentication~\cite{chang_2021} and identification~\cite{discoxyz_2023}. SSI recombined these focal points into a single approach that attributes particular importance to holder control, whereas decentralised identity retained more aspects of Web3 by placing more weight on blockchain-based identity storage. Therefore, all three approaches are considered decentralised, but base their claim of decentralisation on different properties.\\

As a result, the divergence on a definitional and functional level has created insufficiencies in addressing problems both for dApps and real-world applications, which currently cannot be filled by either of the solutions alone. One problem is related to the insufficient provisioning of identification data for previously pseudonymous blockchain applications, such as DAOs. DAOs are ``organization[s] built on smart contracts that can execute autonomously'' without ``central control or management''~\cite{liu_2021}. More concretely, DAOs face the risk of capital depletion through a \textit{sybil} attack, i.e. public keys that mimic properties and behaviours of a unique user, but in fact are doubles~\cite{levine_2005}. In order to decrease the risk of such attacks that means to increase sybil resistance, trustworthy identification data from real-world identities is required in a privacy-preserving format. Neither SSI nor OIDC-based solutions can offer that without significant implementation overhead for dApp developers and DAOs. This has led to a lack of legitimacy of OIDC-based identities and SSI among Web3 practitioners, and a digital identity bridge could fill this gap. The next subsection will describe the technical background of digital identity systems in depth that is needed to conceive a functional digital identity bridge between real-world identities and Web3.

\subsection{The State-of-the-Art in Digital Identity -- Grappling with Fragmentation}\label{sec_2.2}
In this subsection, it is shown that the discourse on the issue of centralisation in digital identity applications and architectures is fragmented. To this end, one of the central arguments is that the dichotomous character of the distinction between (de-)centralised digital identity limits the outcomes of the debate. The issue is further exacerbated because research on comparably centralised digital identity solutions, such as federated identities, is taking place independently from research on decentralised digital identity. Thus, disparate governance and funding structures have led to barriers for collaborative innovation, which is necessary for cross-platform interoperability and standards compatibility. For example, work on OIDC is done under the umbrella of the OpenID Foundation (OIDF)~\cite{oidf_2023}, whereas selective disclosure Javascript Object Notation Web Token (SD-JWT) is standardised at the Internet Engineering Task Force (IETF)~\cite{fett_2023}. Meanwhile, the \emph{Decentralized Identity Foundation} (DIF) focusses on specifying DIDs~\cite{danube_2022}, VCs are standardised at the World Wide Web Consortium (W3C)~\cite{sporny_2023}, and the \textit{Hyperledger} protocols with their anonymous credential specification are controlled by the Linux Foundation through the Trust Over IP Foundation (TOIP)~\cite{hyperledger_2024}. Although SD-JWTs may be structured as VCs and VCs can be exchanged over the updated OIDC-protocols, the lacking cooperation has led to either of the software stacks accounting for frontend-heavy, blockchain-based user applications (dApps). As a result, practitioners have largely avoided Web3 use cases and voiced their concern over the use, development, and contribution of digital identity software code to the relevant governance bodies~\cite{lohkamp_2021}.\\

To cross the chasm between Web3 and the EUDIW beyond eIDAS 2.0, this paper introduces the concept of a digital identity bridge. The notion of bridging is adapted from blockchain interoperability research and refers to the custodial or semi-custodial substitution of one type of cryptographic token through the equivalent of another type of cryptographic token on another distributed database or blockchain~\cite{hardjono_2021}. \textit{Bridging} has gained some traction within the SSI sector. For example, the Hyperledger-based project ValidatedID described the process of onboarding credentials to their electronic ledger-based identity platform~\cite{validatedid_2022}. Bridging relies on the compartmentalisation of identity data through VCs and SD-JWTs, which are submitted to the identity bridge as verifiable presentations (VPs). Furthermore, digital identity bridges use OAuth functionality for retrieving data from a resource server to a client~\cite{hardt_2012}. These two processes can be interpreted as ``spending'' or authorising an identity token. Users then can utilise their cryptographic key pair that are compatible with the Ethereum Virtual Machine (EVM) for encrypting the credential data on the client and upload the structured cipher to the bridging server.\\

Contrary to the assumptions that underpin most generic digital identity solutions on the market, where the correlation of unique verifiable presentations by users is an undesirable privacy risk, in Web3, users desire to build the reputation of their key pair through correlation rather than disclosing their real world identities. Hence, users must rely on a bridging service with pseudonymisation functionalities for using their EUDIW-based credentials in Web3 and benefiting from the user rights under the \emph{EU Directive on the protection of natural persons with regard to the processing of personal data and on the free movement of such data}~\cite{european_parliament_2016}. Even if the bridging service uses a centralised server, the server's signatures over the hashes of the encrypted credentials were considered personal data under the General Data Protection Regulation (GDPR), users should be able to simply delete their credentials making the signatures not refer to any personal data anymore~\cite{kondova_2020}. This does not, however, apply to relying parties as they are currently retaining a copy of the user data with an non-repudiable signature by the holder. For this reason, the proposed digital identity bridge should blind user data where possible and provide users with the option to choose between presenting relying parties with a salted hash of the user data, a predicate, or their data in plain text.\\

Thus, a digital identity bridge fulfils the need of providing measures for \emph{sybil resistance} that are assurances of cryptographic key pairs representing ``a unique human''~\cite{siddarth_2020}. Relying parties follow a risk-based approach in establishing whether a cryptographic key pair represents a unique user and is the only identifier the user controls for the use case. This means that a digital identity bridge must enable relying parties to check pseudonymous identifiers against a second data source, such as an internal reference to credential data provided for other identifiers as plain text or ciphers. By equipping identifiers with additional data in form of a credential, relying parties can increase the difficulty of creating a sybil, which eventually exceeds the utility users receive from the platform. Thereby, the creation of sybils becomes infeasible. Following, most Web3 protocols use probabilistics and game theory for identifying unique pseudonymous users, which distinguishes the approach from holder-binding performed in EUDIWs. The pseudonymity of actors in Web3 is important for explaining the reasons behind the differing approach in Web3. Pseudonymous users utilise Web3 to access use cases that distribute resources or give privileged access to information. The monetary value of these resources leads relying parties to protect the distribution of value by relying on the ``scarcity of humans''~\cite{schaffner_2019} and calculating the probability and cost of a sybil, which is weighed against the value offered. Overall, this approach is similar to strategies for mitigating free riding in general. For this reason the digital identity bridge not only creates references for identity interactions on-chain, but also decouples onboarding transactions from credential presentation and introduces a centralised server for acting as a \textit{zero-knowledge} attestation service. Thereby, the root-of-trust stemming from credentials that originate from an EUDIW are extended into Web3 by the means of attested provenance.\\ 

Therefore, the proposed architecture neither requires the use of an additional identifier that is referenced in a verifiable data registry (VDR) according to the definition in DID-CORE~\cite{sporny_2022}, nor a DID document stored on a VDR. Effectively, the VDR is merged with wallet-like infrastructure making decentralised identity interactions more efficient for users, identity infrastructure providers, and relying parties. The proposed approach, however, is not the first to aim at consolidating established components of decentralised identity infrastructure. On the identifier level, the Key Receipt Infrastructure (KERI) has attempted to introduce key rotation functionality by creating a self-addressing identifier that made DID documents (DDO) and VDR obsolete~\cite{smith_2021}. KERI implemented these identifiers through using a key event log (KEL) that essentially acts as an identifier. KERI identifiers exhibit functionality that can be used for reputation-building because KELs include all actions that have been performed with this identifier~\cite{smith_2021}. Yet, the KEL only includes permutations of the identifier rather than a full log of interactions performed with it.\\

Although a digital identity bridge is expected to transfer the reference of identity data from one identifier to another, as opposed to creating a new type of identifier as KERI does, the two approaches still have one key property in common. KERI specifies the appointment of so-called \emph{witnesses} for conditioning key events, such as key rotation. In the case of the digital identity, the server can be seen as a ``witness'' of the identity data being presented from one identifier and associated with a second one. Identity lifecycles of conventional digital identities that do not use KERI, thus can benefit from increased infrastructure efficiency by reducing the number of context specific issuers needed for providing end-to-end user journeys. Without an identity bridge, issuer services are needed to intermediate each and any domain-specific interaction to verify identity data stemming from another identity domain -- only to issue another set of credentials that are short lived and use case specific.\\

The approach of bridging is more efficient in using existing identity infrastructure and allows using digital credentials for sybil resistance, but it compromises some properties that are common both among centralised, federated, and decentralised digital identity solutions. These tradeoffs are best described as abandoning the paradigm of \emph{self-sovereign identity} (SSI), which has by and large been misunderstood by stakeholders of digital identity providers~\cite{weigl_2022}. It seems that participants in the discourse on the SSI paradigm have assumed that \emph{user control}~\cite{allen_2016} must be implemented through storing credentials on a holder device~\cite{preuschkat_2021b}. Only a few solutions (such as the ``CAS Smart.We'' wallet~\cite{sdika_2020}) provide cloud-based solutions to users. Data privacy and user control of this SSI cloud agent, indeed, is not up to par with other SSI wallets. However, the digital identity bridge juxtaposes the federation-based implementation of \emph{Smart.We} with the learnings of interacting with untrusted servers from the field of cloud security~\cite{dong_2011}.\\

Concerns over user control and user privacy in server-based solutions gave rise to the creation of VCs and anonymous credentials, thus leading to the formulation of the SSI principles by~\cite{allen_2016}. In this light, the proposed digital identity bridge may be evaluated as less stringent in fulfilling the SSI principles \emph{``control''}, \emph{``persistence''}, and \emph{``privacy''}. At the same time, most SSI are not interoperable with identities users already control, such as Web3 identifiers. Only through deriving additional identifiers and corresponding DDOs interoperability is achieved. This requires either compromises by the users, increased identity management overhead for both users and verifiers, or user-flow fragmentation. More concretely, the problem users are facing is binding multiple identifiers to their preferred identifier without having to rely exclusively on SSI wallet infrastructure. An example that illustrates the reasons for the desire to link identifiers independent from the EUDIW and SSI wallets in general, is the proposed solution for bridging EAAs of non-authentic sources into EUDIWs by \emph{Bundesdruckerei GmbH}'s subsidiary ``D-TRUST''~\cite{nguyen_2023}. There, remote signing of EAAs from non-authentic sources by a qualified trust service provider (QTSP) are issued on-demand to EUDIWs, which requires ``central data storage at QTSP''. In other words, by relying on wallet infrastructure -- following the SSI trust triangle~\cite{preuschkat_2021b} -- centralised providers are introduced through the backdoor and still increasing the identity management overhead for users. Rather than QTSPs serving QEAAs from non-authentic sources to users on-demand, the issuance should take place once with user retaining control. Therefore, the digital identity bridge binds identifiers and associated credentials to the primary identifier once, through an attestation on-chain that requires no third-party references of said identifiers or interactive re-issuance process.\\

At the same time, the SSI-centric update to the OIDC protocols focused on interoperability with Web2 and suggested seamless upgradeability of solutions that are currently on the market under OAuth 2.0. In fact, OID4VCI, OID4VP, and SIOP are prescribed by the architecture reference framework (ARF), which specifies the implementation of the EUDIW~\cite{kounis_2023}. This has led to participants in the European discourse on digital identity to conceptualise the implementation of the EUDI as buttons for ``Log-in with EUDIW'', akin to existing OAuth solutions~\cite{padayatti_2023}. Thus, the rise of SSI appears to have redirected the discourse on an updated digital identity framework to square one. None of the solutions developed for the European market are truly interoperable with blockchain networks, hence the lack of addressing challenges for regulatory compliance with the EU Directive on Markets in Crypto-Assets (MiCA)~\cite{european_parliament_2023}, which European operators in Web3 now must face. Moreover, users cannot use their EUDIW for onboarding to a DAO, let alone collecting attestations of their achievements on-chain. \\

Table~\ref{tab_1} summarises the shortcomings of existing digital identity frameworks across Web paradigms and with differing degrees of (de)centralisation. Although this list is not exhaustive, it captures the overall market direction and mainstream discourse on digital identity solutions that are relevant to large-scale private sector initiatives and public sector projects. It is noteworthy that some frameworks and standards are used in conjunction. For example, the OIDC suite and SD-JWTs are used in a vertically integrated fashion for the EUDIW. Others have lost popularity, but are still widely used. PKIs and OAuth can be considered to fall into this category. More broadly, however, this selection of protocols and their shortcomings are intended to substantiate the design choices for the digital identity bridge that was implemented and tested based on the specifications outlined in Section~\ref{sec_3} and following.

\begin{table}[htbp]
\centering
\tiny
\begin{tabular}{p{3cm}p{4cm}p{6cm}}
\hline
\textbf{Technology} & \textbf{Description} & \textbf{Limitations in Web3 Usage} \\ \hline

Public Key Infrastructure (PKI)& A centralised architecture for managing digital certificates through referencing public keys pairs for inducing trust through a certificate authority. & 
\begin{itemize}
    \item Predominantly centralised key storage creates limits user privacy.
    \item Lacking integration with blockchain-based identities
    \item Competing paradigm to Web3 identities
\end{itemize} \\ \hline

Open Authentication (OAuth) & An open standard for access delegation commonly used to grant websites or applications limited access to user information without exposing passwords. & 
\begin{itemize}
    \item Lacks native support for blockchain-based identity verification.
    \item Requires additional setup to integrate with Web3 protocols.
    \item Introduces complexity in managing authentication across Web2 and Web3 systems.
\end{itemize} \\ \hline

OpenID Connect (OIDC) & A Web2-based protocol for user authentication and authorisation extending OAuth 2.0 for standardising user authentication and identity sharing across applications. & 
\begin{itemize}
    \item High reliance on client-server architectures
    \item Limited interoperability with blockchain-based identity solutions
    \item Unsuitable for dApps without user database
\end{itemize} \\ \hline

OpenID Connect for Verifiable Credential Issuance (OIDC4VCI), -- for Verifiable Presentation (OIDC4VP), and Self-Issued OpenID Provider (SIOP) & Update to OIDC that enable issuing and presenting user-controlled credentials. & 
\begin{itemize}
    \item Limited adoption in Web3 due to reliance on Web2 architectures
    \item Challenges privacy-preserving and pseudonymous identity use cases
    \item Nascent standard with high complexity
\end{itemize} \\ \hline

Selective Disclosure JSON Web Token (SD-JWT) & A format for issuing credentials that supports selective disclosure of attributes. & 
\begin{itemize}
    \item Requires additional infrastructure for interoperability with Web3 applications
    \item Current adoption limited to Web2 with privacy trade-offs
    \item Blockchain-based credential systems are not a focus
\end{itemize} \\ \hline

Self-Sovereign Identity (SSI) & A decentralised digital identity framework emphasising user control over identity data through a wallet-based approach. & 
\begin{itemize}
    \item Requires additional infrastructure and SSI wallets for compatibility with Web3
    \item Identity management overhead increases for users and developers
    \item Fragmented implementations lead to inconsistent standards
    \end{itemize} \\ \hline

Key Event Receipt Infrastructure (KERI)& A decentralised framework for managing key rotation and cryptographic identifiers through event logs, eliminating the need for additional components. & 
\begin{itemize}
    \item Lacks built-in mechanisms for linking to other identity frameworks
    \item Niche framework with limited adoption
    \item Web3 interoperability explicitly out-of-scope
\end{itemize} \\ \hline
\end{tabular}
\caption{Overview of Key Digital Identity Technologies and Their Limitations for usage in Web3.}
\label{tab_1}
\end{table}
\FloatBarrier

In sum, the introduction of SSI principles has failed to truly address user needs, who are active participants in Web3. The rigidity of SSI concepts, such as the trust triangle~\cite{preuschkat_2021b}, redirected initiatives for decentralising the EUDI to existing architectures that are used by applications in Web2, as evident by the abandonment of DIDCOMM messaging version 2.1~\cite{curren_2023} as transport protocol for the EUDIW. Finally, the value-laden debate of whether to use blockchain for decentralised identity and the dichotomous distinction between (de-)centralised identities required addressing through an atypical and unconventional identity solution, which is described in detail in the following section.

\section{Technical Design of a Digital Identity Bridge}\label{sec_3}
Previous work, such as on the ``eIDAS bridge'', has shown that the bridging of digital identity data stemming from eIDAS is possible and effective~\cite{burgos_2020}, however, this research on the WIDE digital identity bridge is concerned with the efficiency of the bridging process for Web3 users specifically. Meanwhile, the research on the ``eIDAS bridge'' focused on closed-loop digital identity solutions that are fragmented and directly linked to governmental applications. Although there was some exploration of linking x.509 certificates to DIDs~\cite{bastian_2022} and the conception of a governance framework for DID issuance that relies on secure DNS (DNSSEC), which is called \uppercase{train}~\cite{kubach_2021,martinez_2021}, none of the solutions presented users in Web3 with a viable option for passing sybil resistance checks with their identifier and using their Web3 identifier for submitting credentials to DAOs. The proposed digital identity bridge addresses this limitation through providing two core workflows. In subsection~\ref{sec_3.1} the process of bridging claims is described. It uses the bridging of OAuth payloads as an example. Thereafter, subsections~\ref{sec_3.2} outlines the presentation flow of bridged credentials, which may occur when using a Web3 identifier for accessing a cloud workspace that is permissioned using Web2 identifiers. Finally, the technical architecture enabling both the bridging of claims and their Web3 identifier-based presentation is set forth in subsection~\ref{sec_3.3}.\\

Credential bridging generally consists of two steps. First, the user exports their credential from a legacy identity system. In a second step, credential data is structured and encrypted on users' client devices before being uploaded to the server. The server then logs a signature over the received cipher to a blockchain, referencing it with a hash of the user's EVM-compatible public key. This step can be equated to creating a record of the identity token on the target network and has the following two key benefits. The users delegate the management of their credentials to infrastructure that is more reliable than an edge device, but without trading-off data privacy. Users also can build the reputation of their Web3 address because a track-record of credential usage for the Web3 wallet address is created that does not disclose any personal data.

\subsection{Digital Identity Bridge Architecture Diagram}\label{sec_3.1}
Figure~\ref{fig_1} depicts the architecture of a digital identity bridge, which is designed to bridge the gap between Web3 identity solutions and OIDC protocols used in the context of the EUDIW. The proposed bridging system facilitates the provisioning of user identities from centralised sources for activities in Web3, without the need for additional wallet key and credential management on edge devices. The architecture is clustered in two types of components. The elements in magenta, such as the claims processor and the bridging database, represent key pieces of the identity bridge and fall into the category of core components. Infrastructure that is depicted at the periphery of the diagram mostly belongs to the category of ancillary components, which may change from one implementation of a digital identity bridge to another. The digital identity bridge offers a pragmatic solution for integrating centralised and closed-loop digital identity architectures~\cite{madon_2021} into open-loop and decentralised digital identity networks without trading off privacy.

\subsubsection{Significant Components}\label{sec_3.1.1}
While the previous sections outlined the functionality of the digital identity bridge outlined herein, now, the core components of the identity bridge are presented and defined. The components in this section are denoted as significant because they are under the control and in direct relationship with the solution’s functionality. The following components are essential for processing claims, managing credential storage, and enabling the verification of credentials with centralised provenance on the basis of Web3 identifiers. The below list is intended to add details to the architecture provided in Figure~\ref{fig_1} and clarify the realm that is directly linked to the digital identity bridging system.

	\begin{itemize}
		\item \textbf{Bridging Client}: Processes claims and manages presentation endpoints, utilising Web3 OAuth for authentication.
		\item \textbf{Bridging Server}: Hosts the claims storage, deletion, and fetching modules, protected by Web3 authentication.
		\item \textbf{Bridging Database}: Securely stores encrypted claims that may only be decrypted by the holder, integral to the bridge's data management.
		\item \textbf{Bridging Library}: Provides predefined verification methods for validating the provenance and history of claims. 
		\item \textbf{Identity Logging Smart Contract}: Logs identity interactions on a specific DLT or on multiple DLTs concurrently and exposes it to relying parties.
	\end{itemize}

\subsubsection{Ancillary Components}\label{sec_3.1.2}
In this section, the ancillary components that support the functionality of the identity are described. These components can broadly be sorted into two categories. On one hand, ancillary components are related to offering credential support, thus credential standards and transport protocols. For example,  support for credentials with EUDI provenance and OAuth fall into this category.\\

\begin{figure}[htbp]
\centering{
\includegraphics[scale=0.563]{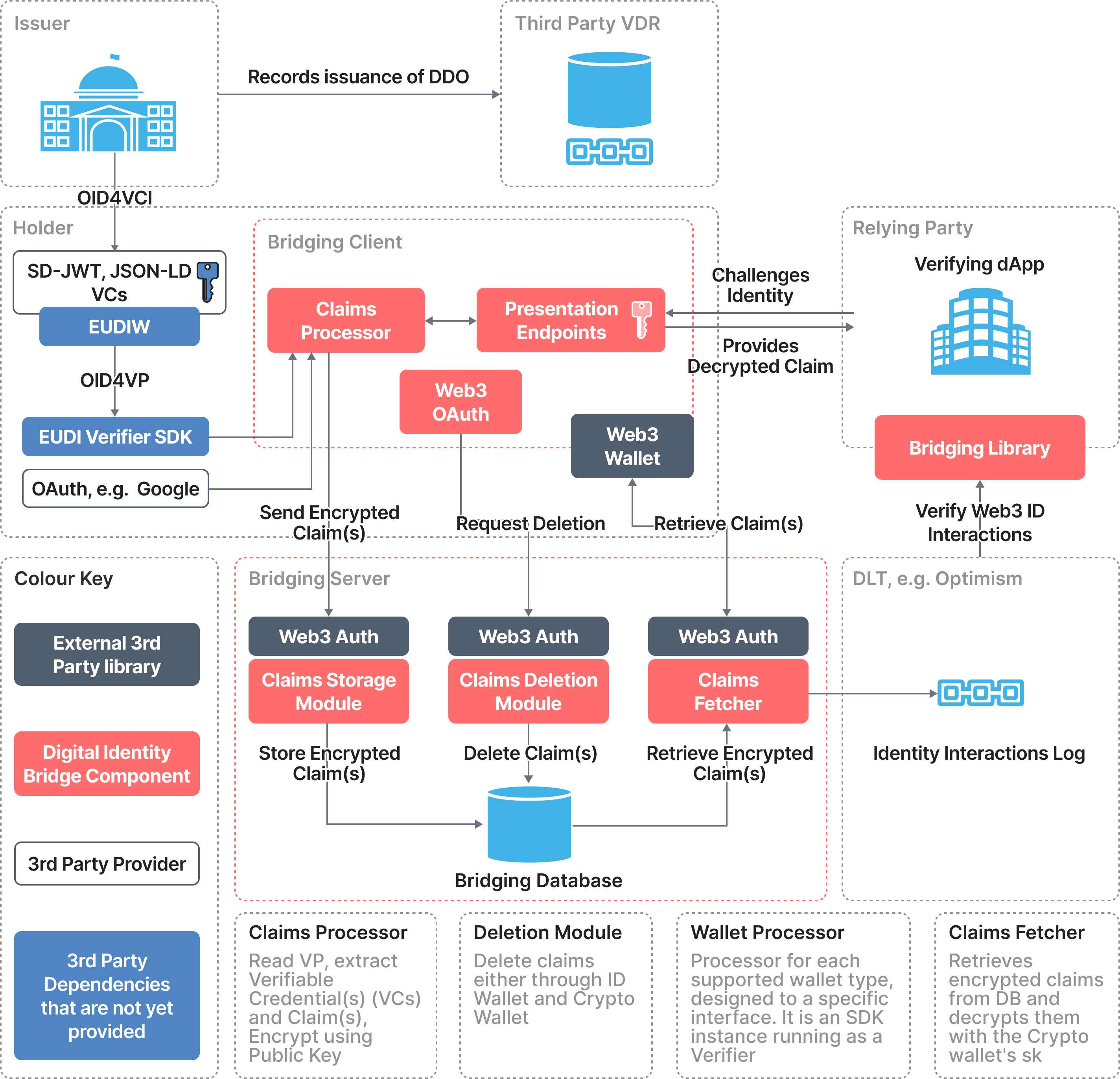}}
\caption{General Architecture of a digital identity bridge on the example of WIDE}\label{fig_1}
\end{figure}

On the other hand, the digital identity bridge has dependencies on the blockchain networks it supports, relying party dApps it integrates with, and VDRs used by issuers and verifiers. By making these dependencies explicit, this section delineates the proposed identity bridging solution from the state-of-the-art, as the novel components and their composition in the system architecture in Figure~\ref{fig_1} is highlighted.

	\begin{itemize}
		\item \textbf{EUDI Verifier SDK}: Provides tools for interacting with the eIDAS 2.0 framework, using OID4VP.
		\item \textbf{Issuer}: Entities issuing digital identities, recording DIDs and VCs issuance, are essential but not part of the bridge.
		\item \textbf{Holder}: Users holding digital identities, managing credentials via an EUDIW and controlling a Web3 wallet.
		\item \textbf{Third Party VDR}: A VDR resolves the DDO of conventional DIDs and operates externally to the bridge.
		\item \textbf{Relying Party}: Utilises the bridging library to verify Web3 identity interactions, interfaces with the bridge externally.
		\item \textbf{External and Third-Party Libraries}: Enhance the bridge's capabilities, representing ongoing development efforts.
		\item \textbf{Distributed Ledger Technology (DLT)}: This example of the identity bridge uses the Optimism EVM for recording identity interactions in smart contracts, but the trust layer depends on changing requirements.
\end{itemize}

\FloatBarrier

\subsection{Bridging claims}\label{sec_3.2}
After the overall architecture of the identity bridging solution was presented in Figure~\ref{fig_1}, this section outlines the steps that are required for providing the functionality of identity bridging. Thus, in the following the process for securely transferring credentials of centralised provenance to Web3 and distributed ledger-backed environments while preserving the privacy of users. The section outlines the system flow for exporting credentials from a centralised provider, encrypted and hashing them before storing them on the bridging server. This explains the core mechanism of the digital identity bridge. Identity bridging involves both client- and server-side processes. More concretely, the claims processor is a client-side application under the control of and executed by Alice. Its trust assumption is equal to the browser wallet \textit{MetaMask}. In a threat scenario, both softwares could be used by an attacker to act counter Alice's intent, when run on a compromised device. The trust assumptions of the claims processor, however, differ significantly from the services provided by the \textit{bridging server}, as the latter is not under Alice's control and should not be trusted with plain user data. The full protocol is described below and assumes steps (1) to (14) are completed in one session that is secured through an \textit{HttpOnly} cookie stored client-side from Alice Sign-in-with-Ethereum~\cite{chang_2021} in step (2).

	\begin{enumerate}[itemsep=3pt]
		\item Alice is connected to an EVM-compatible Web3 wallet address w\textsubscript{1} and connects to the digital identity bridge using her wallet.
		\item The digital identity bridge verifies Alice is the owner of a secret key sk\textsubscript{1}, corresponding to the public encryption key pk\textsubscript{1} and w\textsubscript{1} through signed message verification~\cite{chang_2021}.
		\item Alice selects a credential import i\textsubscript{1} for transferring to w\textsubscript{1} using the digital identity bridge.
		\item Alice exports the pk\textsubscript{1} to the client-side \emph{claims processor}.
		\item Alice uses the client-sided \emph{claims processor} to segregate the issuer data u\textsubscript{1} contained in i\textsubscript{1} from the claims c\textsubscript{1}, ... , c\textsubscript{n} shared by u\textsubscript{1} about Alice.
		\item On Alice's client, the \emph{claims processor} splits c\textsubscript{1}, ... , c\textsubscript{n} into attributes a\textsubscript{1}, ... , a\textsubscript{n} and generates separate payloads for c\textsubscript{1}, ... , c\textsubscript{n} and all attributes a\textsubscript{1}, ... , a\textsubscript{n} for each \emph{c}.
		\item Alice uses the \emph{claims processor} on the client-side to transform all a\textsubscript{1}, ... , a\textsubscript{n} for each and all c\textsubscript{1}, ... , c\textsubscript{n}, into the payloads p\textsubscript{1}, ... , p\textsubscript{n+1}. A claim may contain multiple attributes. The attribute level of data is used to define the number of payloads, whereas payload p\textsubscript{n+1} contains the full set of a\textsubscript{1}, ... , a\textsubscript{n} for each and all c\textsubscript{1}, ... , c\textsubscript{n} to reference the provenance of any a for any c.
		\item Alice encrypts p\textsubscript{1}, ... , p\textsubscript{n+1} using her pk\textsubscript{1}, which gives her the function pk\textsubscript{1}(p\textsubscript{1}, ... , p\textsubscript{n+1}).
		\item Alice hashes p\textsubscript{1}, using SHA-512, giving her h\textsubscript{1}.
		\item Alice presents pk\textsubscript{1}, u\textsubscript{1}, pk\textsubscript{1}(p\textsubscript{1}, ... , p\textsubscript{n+1}), and h\textsubscript{1} to the bridging server.
		\item The bridging server verifies that Alice is authenticated by recovering Alices pk\textsubscript{1} from the signed message in (2) and that the public key matches pk\textsubscript{1} stored in the \textit{HttpOnly} cookie.
		\item The bridging server adds u\textsubscript{1} to the list of issuers Alice has associated with w\textsubscript{1}.
		\item The bridging server stores pk\textsubscript{1}(p\textsubscript{1}, ... , p\textsubscript{n+1}) linked to u\textsubscript{1} for w\textsubscript{1}.
		\item The bridging server signs a dataset d\textsubscript{1} containing w\textsubscript{1} and a hash h\textsubscript{2}. The string h\textsubscript{2} is the result of the SHA-512 function over the encrypted payload of all claims and h\textsubscript{1}, \textit{i.e.} h\textsubscript{2}(pk\textsubscript{1}(p\textsubscript{n+1}), h\textsubscript{1}). The bridging server proceeds to log the signature sig\textsubscript{1} of d\textsubscript{1} and h\textsubscript{2} to the Optimism EVM. So that Alice can prove to have had possession over d\textsubscript{1} at the time of presenting the data to the bridging server.
	\end{enumerate}

\subsection{Presenting claims}\label{sec_3.3}
In this section, the process of presenting identity claims within this digital identity bridge is described. The section details steps users must take for presenting credentials, which were transposed from centralised systems into Web3. The below flow coherently describes both system and user actions in a sequential manner. It commences with relying party \textit{``Bob''} initiating the flow and continues with the holder \textit{``Alice”} decrypting, signing and presenting the credential. The flow concludes with the credential submission and a state change in the smart contract for logging credential interactions.

\begin{enumerate}[itemsep=3pt]
		\item Bob specifies for issuers u\textsubscript{1}, ... , u\textsubscript{n} a dataset d\textsubscript{2}, which specifies a request for plain text claims and blinded claims are needed from Alice.
		\item Bob is sending the presentation configuration d\textsubscript{2} to the bridging server.
		\item Bob's verification website redirects Alice to the digital identity bridge.
		\item  Alice connects with her w\textsubscript{1} to the digital identity bridge and authenticates~\cite{chang_2021}.
		\item The digital identity bridge presents Alice with the request d\textsubscript{2} for c\textsubscript{1}, ... , c\textsubscript{n} and pk\textsubscript{1}(p\textsubscript{1}, ... , p\textsubscript{n+1}) from u\textsubscript{1}.
		\item The digital identity prompts Alice to decrypt pk\textsubscript{1}(p\textsubscript{1}, ... , p\textsubscript{n+1}) using sk\textsubscript{1}.
		\item Alice decrypts pk\textsubscript{1}(p\textsubscript{1}, ... , p\textsubscript{n+1}) using sk\textsubscript{1}.
		\item The digital identity presents Alice with data she is about to present to Bob.
		\item Alice signs a message containing d\textsubscript{2} for her consent to truthfully presenting data to Bob.
		\item The digital identity bridge generates a random token.
		\item The digital identity bridge sends the data that Alice consented to and the token to Bob's server.
		\item The digital identity bridge redirects Alice to Bob with the random token.
		\item The bridging server logs the identity interaction of w\textsubscript{1} with Bob on the Optimism EVM.
\end{enumerate}
\subsection{Sybil Resistance and Trust Assumptions}\label{sec_3.4}
For using the above protocol, the user must access their identity information, using interfaces implemented in the claims processor. The identity information includes the plain text data, their public encryption key, and their wallet address. While users run the claims processor on their client by default, the session for uploading encrypted credentials to the bridging server are secured and time-bound. Thus, in a client-sided setup of the claims processor sophisticated users may alter their claims before encrypting them and uploading them with a valid session to the bridging server. Providing the claims processor as client-side application was chosen to minimise the trust assumptions by the user in the bridging server and maximise user privacy. The architecture design decision is justified by the fact that in Web3-specific cases, it is not in the best interest of the user to alter their data because Web3 applications are pseudonymous by default and establish trust probabilistically~\cite{siddarth_2020}.\\

Instead, for permissioning access to on-chain resources, such as airdrops, applications log on-chain time stamped references for attributes that were possessed by a user before the required attributes for accessing a resource were known. In other words, a contribution to a GitHub repository is only publicised as a requirement for receiving an airdrop after a snapshot of on-chain references for of this work was taken. Thus, by decoupling on-chain identity attribute creation at time t\textsubscript{0} from publishing the on-chain identity requirement at t\textsubscript{n} and verifying the presence of a corresponding identity attribute on chain at t\textsubscript{n+1} reduces the risk of identity gaming by the any increase of the distance between t\textsubscript{n} and t\textsubscript{n+1}. In other words, the cost of creating a fraudulent sybil identity increases the longer identity creation and identity verification are apart, as well as with an increasing number of unrelated uses of the identity in between~\cite{siddarth_2020}. Furthermore, any tampering with the user data would increase the entropy of data being logged on chain, further reducing the chance of a user possessing the right attribute.\\

As every use of a given claim through the identity bridge with any dApp is logged on-chain, additional trust requirements for an identity can be introduced by a verifier, such as the user having utilised their bridged identity n times at m different dApps. In this regard, on-chain logging documented under step (14) in Section~\ref{sec_3.2} and step (13) in Section~\ref{sec_3.3} is not programmatically included in the protocol, because stakeholder feedback collected from RaidGuild DAO and Web3 industry participants highlighted the preference of verifiers to define datasets and build verification dApps on a case-by-case basis without being locked into any given identity framework. In other words, the unspecified use of on-chain logs in the WIDE bridging service are a direct response to the limitations of existing digital identity frameworks, outlined in Table~\ref{tab_1} and highlight its flexibility.\\

In case, these trust assumptions and a probabilistic threat model is insufficient for verifiers. Verifiers can request the claims processor to be run on a second server separate to the bridging server, attesting to the correctness of the identity data in plain text. This would increase the trust requirements in the bridging solution by users and reduce their privacy, but provide additional safeguards against the tampering of data before encryption and upload. Users would be free to choose between the two onboarding mechanisms, maintaining control and data sovereignty. While this is a possible enhancement of the solution, it is not yet implemented and still relies on trust in a pseudonymous asymmetric key pair the identity data is associated, encrypted, and presented with. Therefore, this altered solution achieves secure and privacy-preserving transfer of credentials of centralised provenance to Web3 and distributed ledger-backed environments, enabling probabilistic sybil resistance by combining:

\begin{enumerate}[itemsep=3pt]
		\item the requirement for users to authenticate,
		\item proving the possession of the private key for the public encryption key and address,
		\item on-chain logging of identity uploads and presentations,
		\item presentation signing, and,
		\item out-of-band verification of on-chain activity for any given wallet address using WIDE.
\end{enumerate}

\section{Application of WIDE}\label{sec_4}
\begin{figure}[htbp]
        \centering
        \includegraphics[scale=0.17]{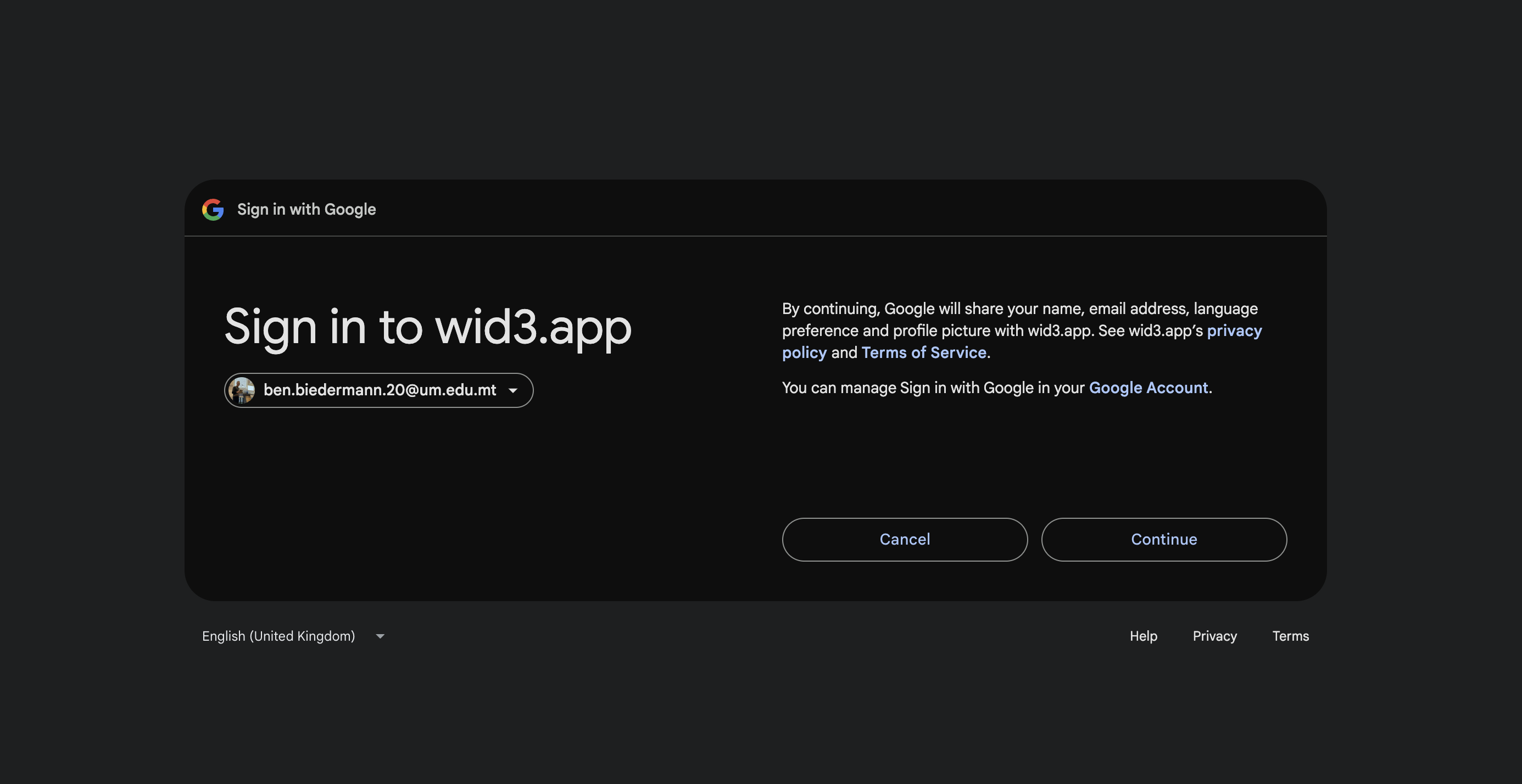}
        \caption{OIDC Data Export}
        \label{fig_2.1}
    \end{figure}

\begin{figure}[htbp]
        \centering
        \includegraphics[scale=0.2]{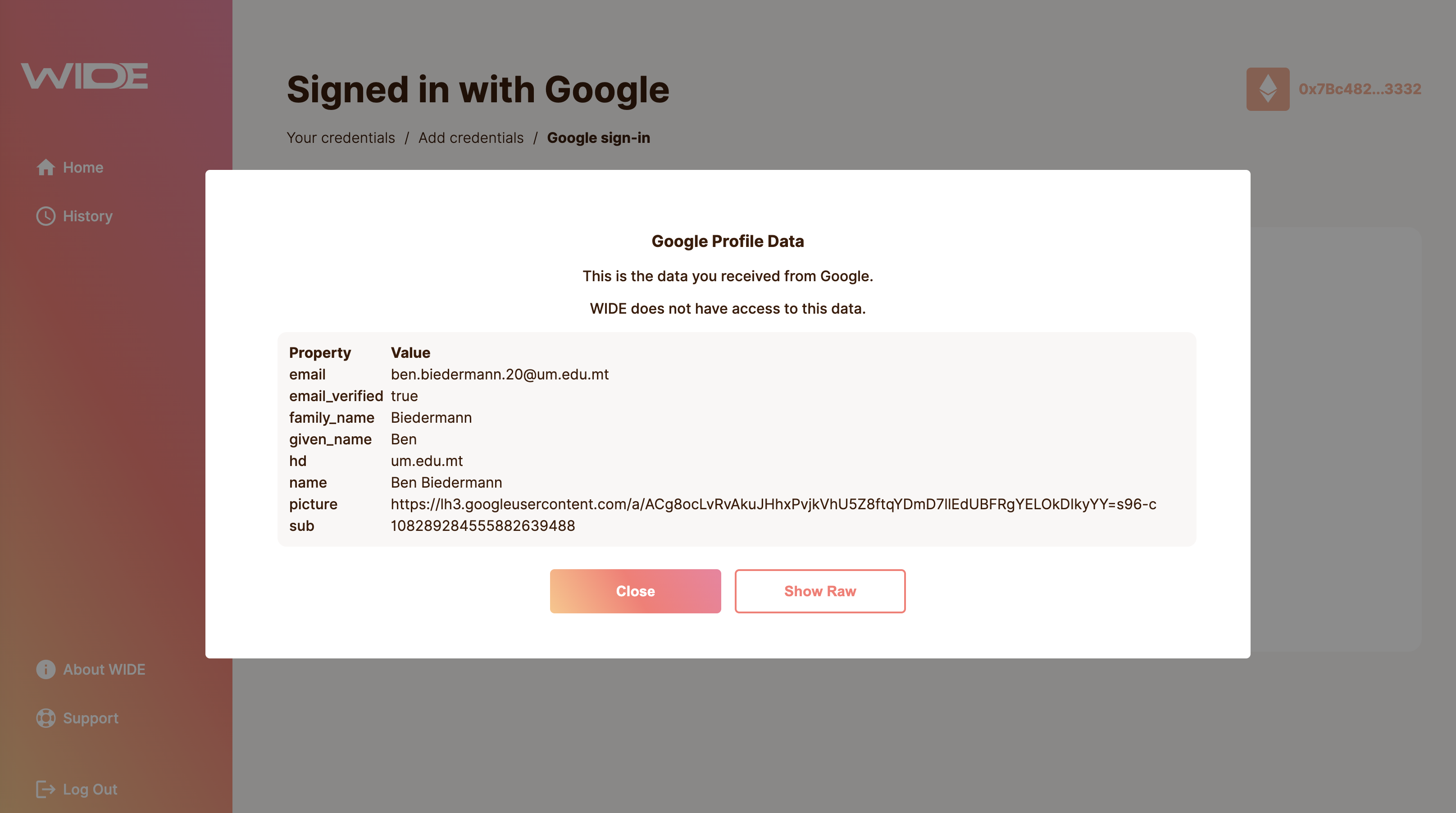}
        \caption{Plain Text Data Review}
        \label{fig_2.2}
    \end{figure}

\begin{figure}[htbp]
        \centering
        \includegraphics[scale=0.2]{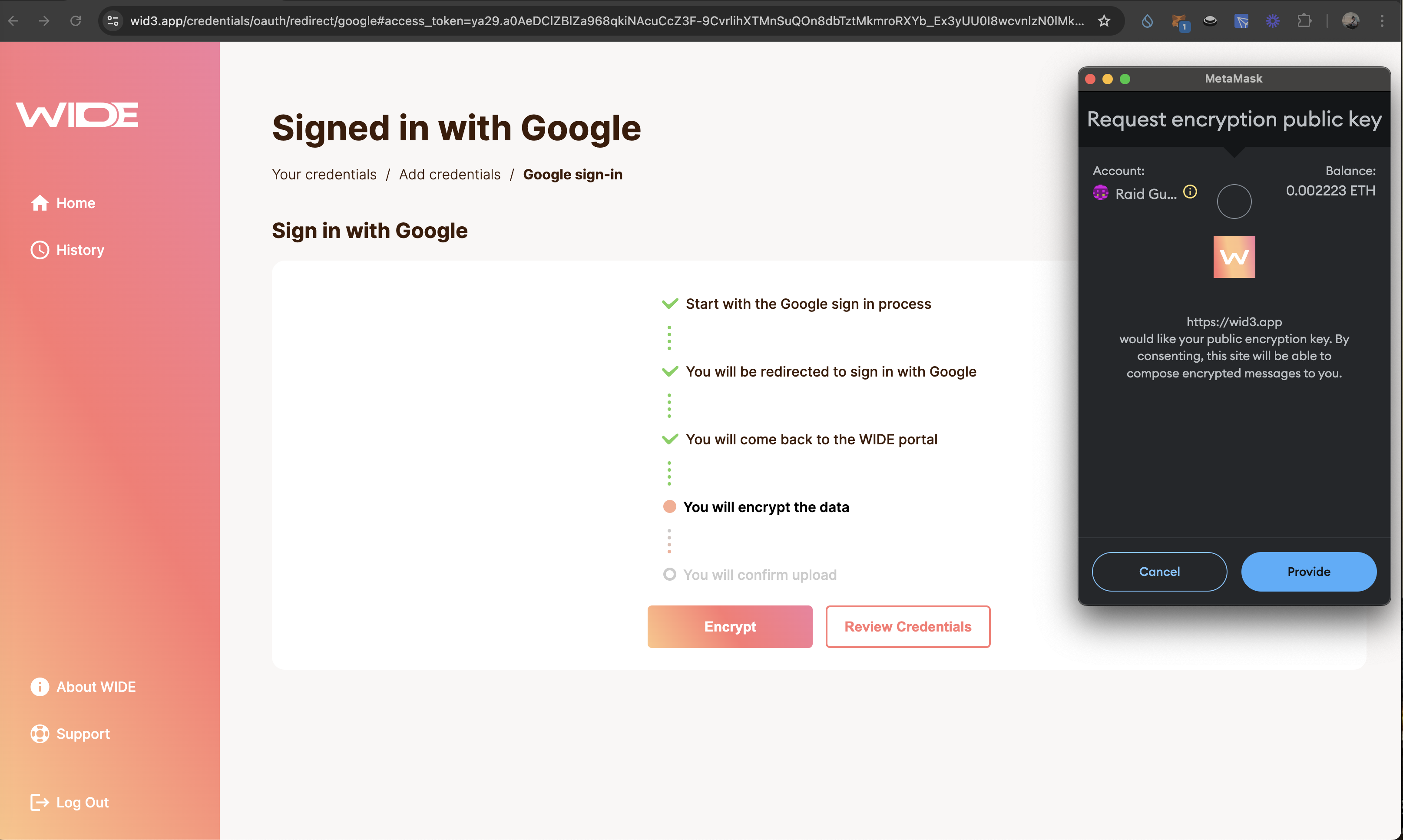}
        \caption{Data Encryption}
        \label{fig_2.3}
    \end{figure}

Based on the protocol outlined in previous sections a proof-of-concept was built, implementing the digital identity bridging protocol. The proof-of-concept served the use case of recent graduates exporting their OIDC-based credential of university affiliation for using it in a Web3 customer relation management tool, where users received credentials showcasing their current projects, skills, and contact information. The use case implementation served for verifying the applicability of the concept for a digital identity bridge in Web3. In collaboration with a DAO the researchers evaluated the identity and data management needs of the DAO’s members, as well as analysed the state of credential management on an organisational level. The application for bridging credentials \textit{WIDE} is accessible at \href{https://wid3.app/}{wid3.app} and the verifier application for this use case is available at \href{https://dungeonmaster.wid3-demo.app/}{dungeonmaster.wid3-demo.app}. Figures~\ref{fig_2.1} to~\ref{fig_2.3} display the onboarding flow in an abridged manner, focusing on the technically relevant and innovative steps.

\FloatBarrier

\begin{figure}[htbp]
   \centering
    \begin{minipage}[htbp]{0.45\textwidth}
        \centering
        \includegraphics[scale=0.08]{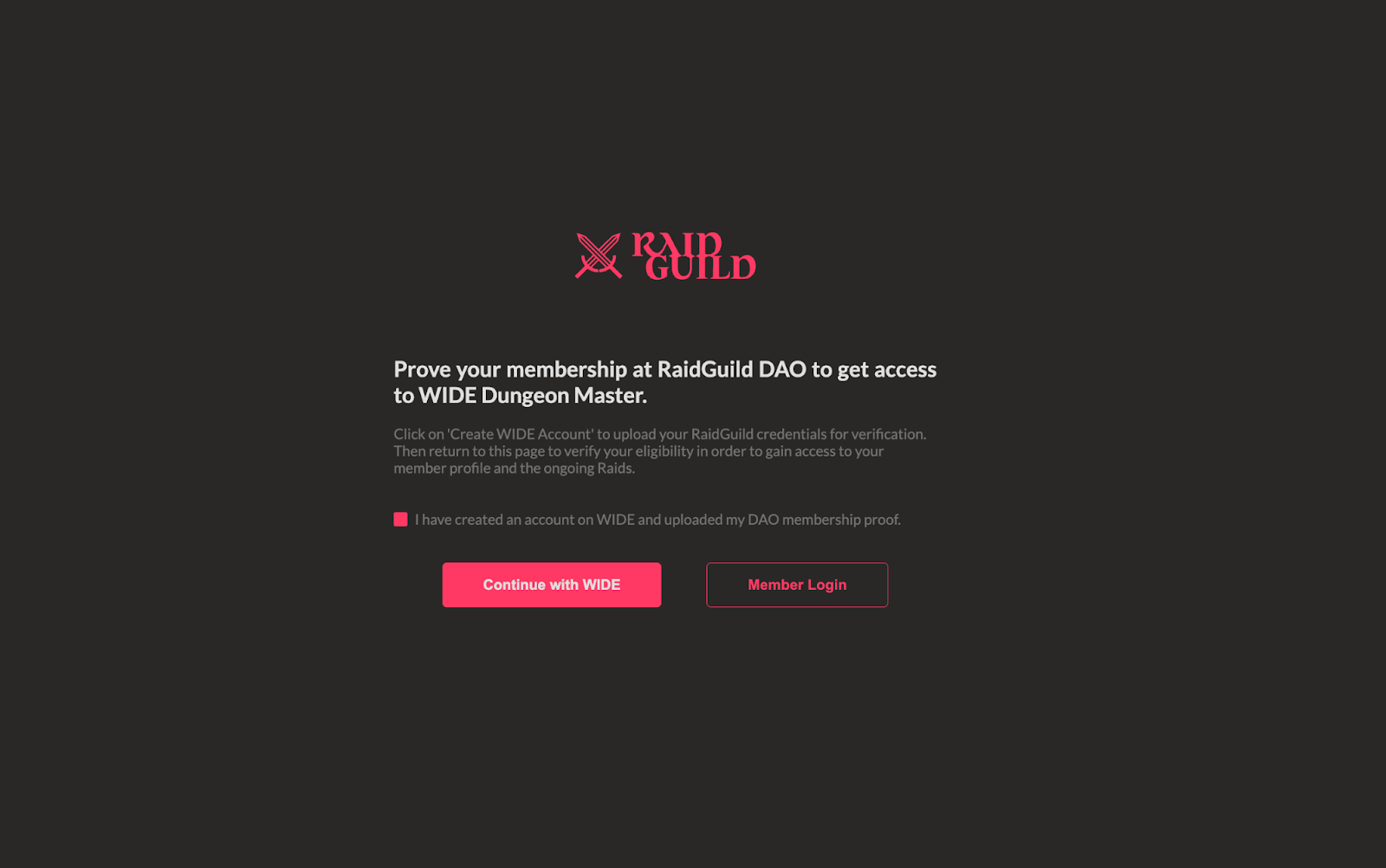}
        \caption{Log-in with WIDE at Verifier}
        \label{fig_3.1}
    \end{minipage}
    \begin{minipage}[htbp]{0.45\textwidth}
        \centering
        \includegraphics[scale=0.17]{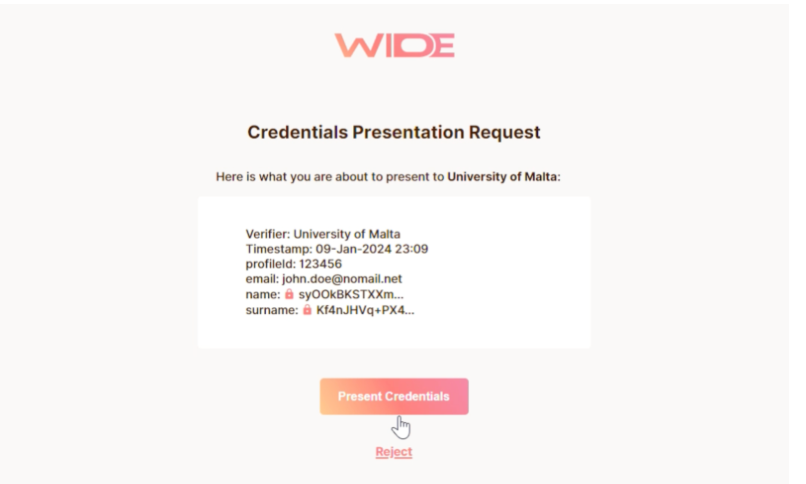}
        \caption{Presenting Blinded and Plain Text Data}
        \label{fig_3.2}
    \end{minipage}
\end{figure}

The DAO that participated in the research is ``RaidGuild'', which is a collective of technical freelancers servicing core Web3 infrastructure and applications, such as decentralised social platforms, decentralised finance (DeFi) protocols, and on-chain attestation services. Quickly it became apparent that the ``RaidGuild'' DAO is a special case, because of their focus on technology services. The abundance of developer resources in the DAO has led to the creation of a plethora of custom tools and processes for all aspects of their operation, which is not usually the case in any organisation, not just in Web3. When evaluating the existing toolkit of ``RaidGuild'' more closely, however, it became apparent that despite their custom tooling, membership data was fragmented across different (de-)centralised databases. Hence, members may have a dedicated environment for managing all their credentials, but the credential data of individual members changed so quickly that the data on their membership data management platform was almost always out-of-date. This gave rise to the opportunity to test the digital identity bridge specifically for the maintenance of up-to-date membership data through continuously aggregating various professional credentials and providing the latest attested copy during every sign-in process. The user experience for this WIDE functionality related to the protocol under Section~\ref{sec_3.3} is displayed in Figures~\ref{fig_3.1} and~\ref{fig_3.2}.

\FloatBarrier

\begin{figure}[htbp]
    \begin{minipage}[htbp]{0.45\textwidth}
        \centering
        \includegraphics[scale=0.11]{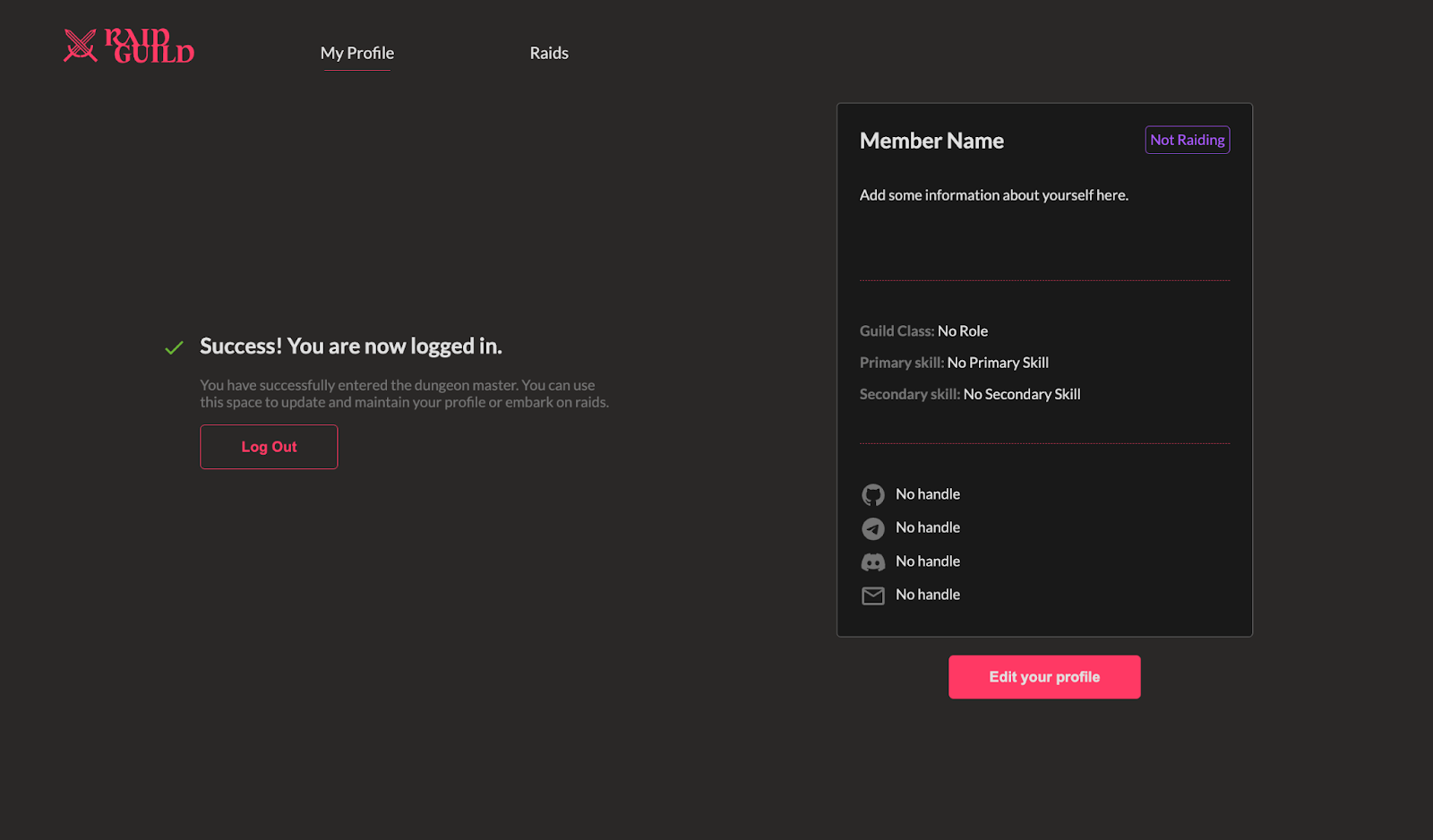}
        \caption{WIDE-enabled RaidGuild Member Profile}
        \label{fig_4.1}
    \end{minipage}
    \begin{minipage}[htbp]{0.45\textwidth}
        \centering
        \includegraphics[scale=0.1]{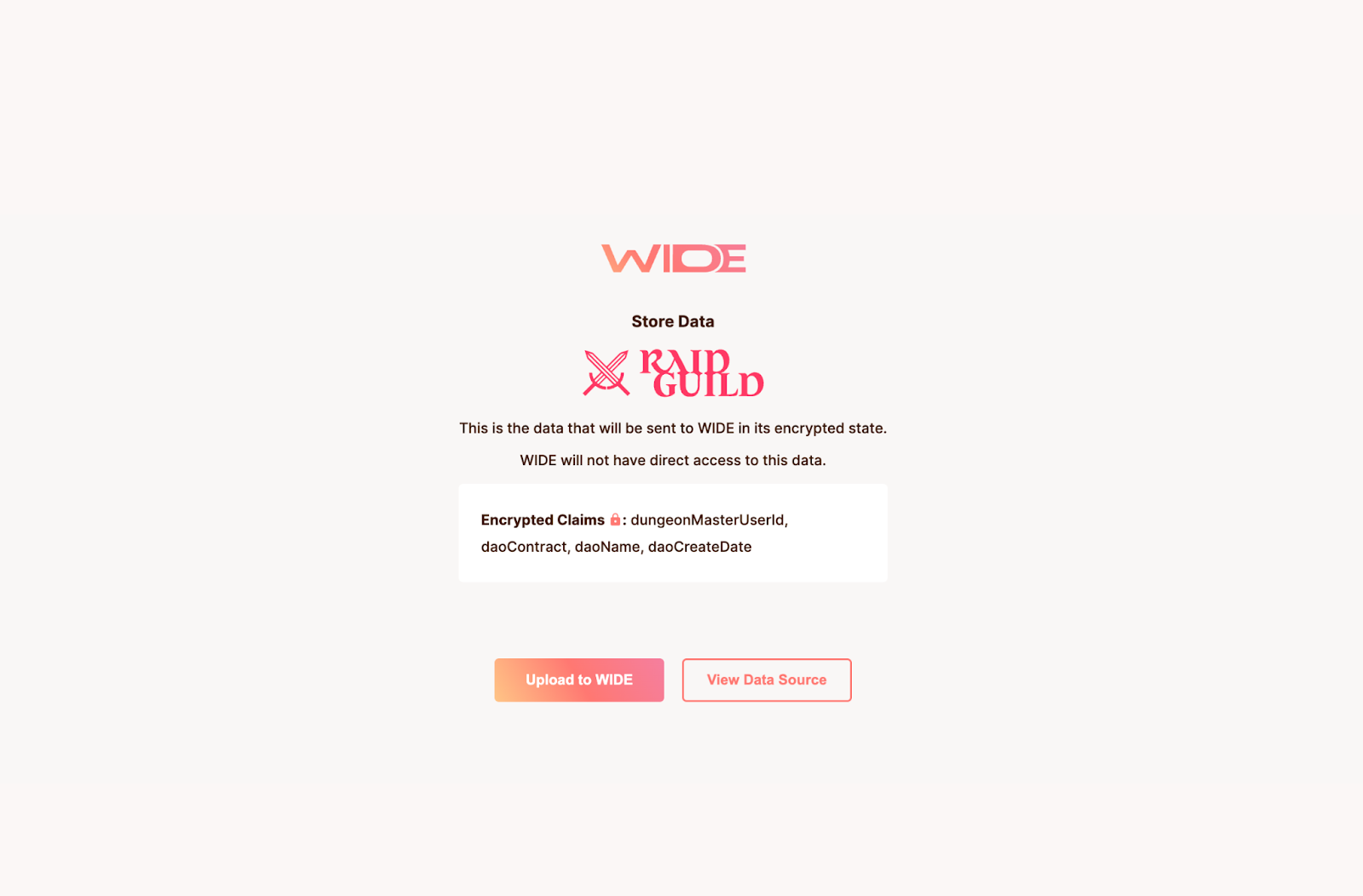}
        \caption{WIDE-export of Member Profile}
        \label{fig_4.2}
    \end{minipage}
\end{figure}

More concretely, users were able to present their university credentials to onboard to RaidGuild customer relation management tool, where they joined projects and showcased their skills. Once signed up, users received a credential from the RaidGuild DAO verifier, which issued user the data they have added to the RaidGuild customer relations management dApp back to their WIDE-stored identity. As a result, users were able to tie their otherwise off-chain identity usage and acquisition of skills to their Ethereum addresses, creating trust in their pseudonymous identifier and a track record for a contributor to a DAO. At RaidGuild, this reputation serves as a means to select contributors, who are conferred governance rights on the DAO's smart contracts and are eligible for permissionless reimbursements for completed work that was undertaken internally. From the below Figures~\ref{fig_4.1} and~\ref{fig_4.2} it becomes apparent to what extent the WIDE solution expands the features of existing Web2 to solutions in the realm of Web3. Now, user not only can change, update, and \textit{own} their identities, they also collect off-chain credentials for their regular work that is still useful for verifiers relying only on crypto-economic systems on blockchain networks. That is because all interactions displayed from Figure~\ref{fig_2.1} to Figure~\ref{fig_4.2} are referenced on-chain.

\FloatBarrier

In this regard, maintaining updated records of members' expertise and experience, appeared to be especially important to the DAO because it provides expert technical services to Web3 organisations~\cite{raidguild_2024}, which require verified professional credentials. The DAO must vet the skills and certifications of new members. Without an identity bridge, it verifies the on-chain activity of applicants' wallet addresses and relies solely on their self-attested skills, who provide references to GitHub repositories and other Web2 accounts for supporting their claims. Meanwhile, European higher education institutions have already commenced issuing electronically signed university diplomas in a machine-readable format~\cite{cedefop_2015}. The DAO can benefit from such official certification through their self-asserting presentation with a Web3 wallet address. Even though it does not directly receive the credential data cryptographically verifiable from the university issuer, having the diploma metadata available is a significant improvement. Irrespective of the theoretical technical capabilities of decentralised identity systems, the vetting of real-world identities for uses other than regulatory compliance is met by DAO members with suspicion and dissent. Thus, the digital identity bridge offers a compromise between the technological capabilities of the eIDAS 2.0 framework and market needs in Web3.\\

Overall, the digital identity bridge is designed for striking a compromise, which is evident when turning to the identity and data management challenges DAOs and their members face. Web3 contributors must deal with manifold Web2 and Web3 identifiers across systems that are early stage or experimental, as well as in established cloud environments, such as \emph{Google Workspace}. In other words, yet another identifier and wallet for managing the data associated with it, is not solving the needs of DAO members. Instead, it is more useful to trade-off root trust for a single (data)space where individuals can aggregate identities from different sources, manage the identity data themselves without the need for installing an application on their devices, and having transparency over their identity use. In sum, the provision of an attribute referencing the provenance of credentials, while focusing on the aggregation of disparate identities and the creation of on-chain records, makes the digital identity bridge useful for DAOs and Web3 at large. This responds to the initial research question by emphasising the positive and desired correlation effects of long-lived pseudonymous identifiers that are used for reputation building and digital identity data management.

\section{Conclusions}\label{sec_5}
The proposed digital identity bridge may have struck a compromise for enabling the use of both centralised and decentralised identities Web3 identifiers, however, more work is needed for integrating OID4VC-based SD-JWTs originating from a EUDIW in compliance with eIDAS 2.0. Furthermore, the digital identity bridge only bridges in one way, that is, from established digital identity systems to Web3. In the future, the bridge must also allow for Web3 identities and data to be bridged into existing digital identity frameworks. Meanwhile, the centralised bridging server may give rise to critique by maximalist voices advocating for decentralisation. The reconciliation of scepticism and compliance with eIDAS will remain a challenge for any bridging architecture and service, however, the exploration of decentralised governance mechanisms of the infrastructure is a promising next step for maturing the concept. Such research can also inform work on the business model and sustainability of a digital identity bridge, which will incur costs both for compliance with eIDAS 2.0 and by logging claims on a blockchain network. Advanced cryptographic methods and tools, such as cryptographic accumulators may offer cost-reducing benefits while enhancing the trust in the system. Lastly, the digital identity bridge has not been integrated or tested with governmental digital identity frameworks outside of the EU, research on the integration of the Indian \emph{Aadhaar} system or electronic residence schemes in small jurisdictions are an anticipated research avenue~\cite{tammpuu_2022}.\\

In conclusion, the proposed system for bridging digital identities from established identity frameworks into Web3 proves the concept of digital identity bridges. It was shown how this architecture transforms closed-loop identity systems into network-based identities that follow an open loop logic. Today, this system can already lessen the burden of DAO members and juxtaposes economic identities that are based on on-chain financial transaction volume, such as \emph{GitCoin Passport}~\cite{gitcoin_2024}. Thus, this work is intended at bringing both voices from opposite sides of the discourse on (de)-centralised digital identity discourse closer together.
\section*{Acknowledgements}\label{ack}
This work has been funded by the EU in the framework of the NGI TRUSTCHAIN project. TRUSTCHAIN is a European project funded by the European Commission under HORIZON-CL4-2022-HUMAN-01-03 and funded under the grant number: \href{https://cordis.europa.eu/project/id/101093274}{101093274}. 

\printbibliography

\end{document}